\documentclass[conference]{IEEEtran}
\IEEEoverridecommandlockouts

\usepackage{cite}
\usepackage{amsmath,amssymb,amsfonts}
\usepackage[acronym]{glossaries}
\usepackage{algorithmic}
\usepackage{graphicx}
\usepackage{textcomp}
\usepackage[table]{xcolor}  
\usepackage{booktabs, makecell, booktabs,multirow} 
\usepackage{url} 
\usepackage{hyperref}
\usepackage[capitalise]{cleveref}
\usepackage{subcaption}
\usepackage{flushend}
\usepackage{svg}

\def\BibTeX{{\rm B\kern-.05em{\sc i\kern-.025em b}\kern-.08em
    T\kern-.1667em\lower.7ex\hbox{E}\kern-.125emX}}
    
\newacronym{sol}{SOL}{String Object List}
\newacronym{nlp}{NLP}{Natural Language Processing}
\newacronym{gnn}{GNN}{Graph Neural Network}
\newacronym{gan}{GAN}{Graph Analysis Network}
\newacronym{gatn}{GAtN}{Graph Attention Network}
\newacronym{sa}{SA}{Self Attention}
\newacronym{msa}{MSA}{Multihead Self Attention}
\newacronym{ga}{GA}{Graph Attention}
\newacronym{ssl}{SSL}{Self-Supervised Learning}
\newacronym{kl}{KL}{Kullback-Leibler}

\newcommand{\methodname}{DSVAE}
\newcommand{\methodfullname}{Disentangled Spectrogram Variational Auto Encoder (DSVAE)}

\newacronym{qmf}{QMF}{Quadrature Mirror Filter}
\newacronym{mdct}{MDCT}{Modified Discrete Cosine Transform}
\newacronym{imdct}{IMDCT}{Inverse Modified Discrete Cosine Transform}
\newacronym{mfcc}{MFCC}{Mel-Frequency Cepstrum Coefficients}
\newacronym{cqt}{CQT}{Constant-Q Transform}
\newacronym{stft}{STFT}{Short Time Fourier Transform}
\newacronym{dft}{DFT}{Discrete Fourier Transform}
\newacronym{dct}{DCT}{Discrete Cosine Transform}
\newacronym{fft}{FFT}{Fast Fourier Transform}
\newacronym{auc}{AUC}{Area Under Curve}
\newacronym{roc}{ROC}{Receiver Operating Characteristics}
\newacronym{eer}{EER}{Equal Error Rate}
\newacronym{lfccs}{LFCCs}{Linear Frequency Cepstral Coefficients}
\newacronym{mfccs}{MFCCs}{Mel Frequency Cepstral Coefficients}
\newacronym{cqccs}{CQCCs}{Constant Q Cepstral Coefficients}
\newacronym{cfccs}{CFCCs}{Cochlear Filter Cepstral Coefficients}
\newacronym{svm}{SVM}{Support Vector Machine}
\newacronym{gmm}{GMM}{Gaussian Mixture Model}
\newacronym{tssdnet}{TSSDNet}{Time-Domain Synthetic Speech  Detection Net}
\newacronym{aac}{AAC}{Advanced Audio Coding}
\newacronym{flac}{FLAC}{Free Lossless Audio Codec}
\newacronym{mlp}{MLP}{Multi Layer Perceptron Network}
\newacronym{bce}{BCE}{Binary Cross Entropy}
\newacronym{auprc}{AUPRC}{Area Under Precision Recall Curve}
\newacronym{rocauc}{ROC-AUC}{Area Under the Receiver Operating Characteristic Curve}
\newacronym{cnn}{CNN}{Convolutional Neural Network}
\newacronym{ssast}{SSAST}{Self-Supervised Audio Spectrogram Transformer}
\newacronym{passt}{PaSST}{Patchout faSt Spectrogram Transformer}
\newacronym{cct}{CCT}{Compact Convolution Transformer}
\newacronym{fnr}{FNR}{False Negative Rate}
\newacronym{fpr}{FPR}{False Positive Rate}
\newacronym{rnn}{RNN}{Recurrent Neural Network}
\newacronym{cqost}{CQ-OST}{Constant-Q Octave Subband Transform}
\newacronym{icqcc}{ICQCC}{Inverted Constant-Q Cepstral Coefficients}
\newacronym{dnn}{DNN}{Deep Neural Network}
\newacronym{se}{SE}{Squeeze-and-Excitation}

\glsdisablehyper

\newcommand{\etal}{\textit{et al}.\@ }
\newcommand{\ie}{\textit{i.e.},\@ }
\newcommand{\eg}{\textit{e.g.},\@ }

\newcolumntype{L}[1]{>{\raggedright\let\newline\\\arraybackslash\hspace{0pt}}m{#1}}
\newcolumntype{C}[1]{>{\centering\let\newline\\\arraybackslash\hspace{0pt}}m{#1}}
\newcolumntype{R}[1]{>{\raggedleft\let\newline\\\arraybackslash\hspace{0pt}}m{#1}}

\definecolor{kellygreen}{rgb}{0.3, 0.73, 0.09}
\definecolor{amber}{rgb}{1.0, 0.75, 0.0}
\definecolor{amethyst}{rgb}{0.6, 0.4, 0.8}

\newcommand{\td}{$^\dagger$}
\newcommand{\tdd}{$^\ddagger$}
\title{\methodname: Interpretable Disentangled Representation for Synthetic Speech Detection}
\author{
\parbox{0.95\linewidth}{\centering Amit Kumar Singh Yadav\td \hspace{2em} Kratika Bhagtani\td \hspace{2em} Ziyue Xiang\td \hspace{2em}  Paolo Bestagini\tdd \\
\centering Stefano Tubaro\tdd \hspace{2em} Edward J. Delp\td
    \vspace*{0.5em}\\
    \small\centering \td  Video and Image Processing Lab (VIPER), School of Electrical and Computer Engineering, \\
    \small\centering  Purdue University, West Lafayette, Indiana, USA\\
    \small\centering \tdd Dipartimento di Elettronica, Informazione e Bioingegneria, Politecnico di Milano, Milano, Italy
}
}

\begin{document}
\maketitle
\thispagestyle{plain}
\pagestyle{plain}
\begin{abstract}
Tools to generate high quality synthetic speech that is perceptually indistinguishable from speech recorded from human speakers are easily available.
Many incidents report misuse of synthetic speech for spreading misinformation and committing financial fraud.
Several approaches have been proposed for detecting synthetic speech.
Many of these approaches use deep learning methods without providing reasoning for the decisions they make.
This limits the explainability of these approaches.
In this paper, we use disentangled representation learning for developing a synthetic speech detector. 
We propose \methodfullname~ 
which is a two staged trained variational autoencoder that processes spectrograms of speech to generate features that disentangle synthetic and bona fide speech.
We evaluated \methodname~ using the ASVspoof2019 dataset.
Our experimental results show high accuracy ($>98\%$) on detecting synthetic speech from 6 known and 10 out of 11 unknown speech synthesizers.
Further, the visualization of disentangled features obtained from ~\methodname~ provides reasoning behind the working principle of \methodname~, improving explainability of our method.
\methodname~ performs well
compared to 
several existing 
methods.
Additionally, \methodname~ works 
in practical scenarios such as detecting synthetic speech uploaded on social platforms and against simple attacks such as removing silence regions.
\end{abstract}


\glsresetall
\section{Introduction}

Generating perceptually human-like synthetic speech has been of interest for a long time.
Traditional synthetic speech methods either use source modeling or simple waveform cut-paste techniques to generate synthetic speech~\cite{klatt2016,zakariah2017digital, bhagtani2022overview}. 
Recent deep learning methods can generate high-quality and semantically consistent long-duration speech signals~\cite{kim2021conditional, fastspeech_2_iclr_2021, popov2021gradtts}.
Some deep learning methods can generate synthetic speech that can mimic language accents and impersonate persons using 3 seconds of their speech~\cite{wellsaid,vall_e, gan_voice_impersonation_icassp}.
The diminishing perceptual difference between synthetic speech and bona fide human speech is useful in applications such as voice assistants, games, and e-learning~\cite{hoy_2018, microsoft_2021, disney_movie}.
However, several incidents have reported misuse of synthetic speech for spreading misinformation~\cite{allyn_2022}, committing financial fraud~\cite{smith_2021}, and using for impersonation attacks~\cite{gan_voice_impersonation_icassp}.
To prevent such misuse of synthetic speech, there is a need for development of methods that can detect synthetic speech.

Several methods have been proposed for detecting synthetic speech~\cite{tssdnet_2021,subband_cqcc_mlp_19, eightfeatures_mlp_21,spec_cqcc_resnet_se_27,spec_vgg_sincnet_28, fft_cnn_29,li2021replay}.
These methods have shown promising detection accuracy but lack in interpretability/explainability.
By interpretability we mean that the method provides reasoning for the decision it makes~\cite{explainable_ai_2022}.
 Kien~\etal~ argued that representations produced by a neural network are interpretable by humans only if they are discriminative w.r.t one attribute (characteristic) ~\cite{disentagled_representation}.
For example, in speech synthesis, if one of the representation controls only the accent of the speech signal and a different instance of this representation generates a different accent in speech, then the representation is discriminative w.r.t only one attribute \ie~accent.

Disentangled representation learning has been proposed as an approach to learn interpretable representations from a neural network~\cite{Zhu_2021_CVPR, wacv_disentangled_2022,eccv_disentagled_2020, eccv_disentagled_2020_generic}. 
This approach separates the network's latent representations into different components.
Tang~\etal~used disentangled representation learning to control certain regions while generating synthetic images~\cite{icpr_2021_image_generation}.
Disentangled representation learning has also been used to generate interpretable representations for face anti-spoofing~\cite{eccv_disentagled_2020_generic, eccv_disentagled_2020}.
The disentangled representation learnt for face anti-spoofing will be independent of the content in image and will be different depending on only one attribute \ie~whether face image is spoofed or is pristine.
For example, 
two images with identical faces/content in them will have different/farther representations if one image is spoofed and other image is pristine.
So, visualization of the disentangled representation will show evident discrimination between representations of pristine and spoofed face images.
Such a visualization can provide forensic analysts with reasoning behind working of the detector making it more explainable.
Disentangling the representation improves the method's ability to generalize to unseen face-spoofing attacks~\cite{wacv_disentangled_2022, eccv_disentagled_2020_generic}.

Generalization to unseen attacks is also important for synthetic speech detectors. 
One major challenge involved in synthetic speech detection is the increasing number of synthesizers which makes it practically infeasible to include synthetic speech from all possible speech synthesizers during training. 
This demands that synthetic speech detection methods should generalize to 
unknown synthesizers that were not included in training.
Motivated by the success of disentangled representation learning in face anti-spoofing~\cite{wacv_disentangled_2022, eccv_disentagled_2020_generic, eccv_disentagled_2020, Zhu_2021_CVPR}, controlled image generation~\cite{icpr_2021_image_generation}, voice conversion~\cite{icassp_voice_style_transfer, acm_sigsac_2020}, and speech generation tasks~\cite{icassp_2020_voice_generation}, we propose a method to use it for developing a
detector 
for synthetic speech detection.



The contributions of our paper are as follows.
(a) We propose \methodfullname, a dual stage network using speech spectrogram for synthetic speech detection. (b) The disentangled representations obtained from 
DSVAE when visualized provide explanation behind the decision made by the detector and enable generalization of our method to 
10/11 unknown synthesizers.
(c) We use disentangled representation and generate an activation map to highlight regions in the spectrogram that help to make the decision.
(d) \methodname~ performs better than several existing methods using spectrogram 
and baselines provided in the ASVspoof2019 Challenge. 
(e) We also investigate performance of \methodname~ in two practical scenarios: when synthetic speech is uploaded on social platforms and during simple attacks like silence removal from speech.
\methodname~ performs well in these scenarios.

The rest of the paper is organized as follows. 
In \cref{sec:related-work} we discuss existing methods for synthetic speech detection and disentangled representation learning.
~\cref{sec:method} describes  \methodfullname. 
\cref{sec:setup} describes the dataset used in our experiments and the implementation details. \cref{sec:results} discusses our experimental results.
Finally, \cref{sec:conclusions} concludes the paper with our last remarks.
\section{Related Work}\label{sec:related-work}
In this section we describe existing work in synthetic speech detection and disentangled representation learning.
\subsection{Synthetic Speech Detection}
Some methods for synthetic speech detection use hand crafted features such as cepstral coefficients to detect synthetic speech~\cite{asvspoof19, li2021replay, cqccs, mfccs,albadawy2019detecting}. 
These hand crafted features include temporal and spectral features such as \gls{cqccs}~\cite{cqccs}, \gls{mfccs}~\cite{mfccs}, \gls{cqt}~\cite{li2021replay}, and \gls{lfccs}~\cite{li2021replay}.
Since feature selection procedure can be tedious, other approaches process the time-domain speech signal as a sequence~\cite{tssdnet_2021} using \gls{rnn}~\cite{he2016deep}.
The Fourier Transform can be used to convert a time domain speech signal into an image representation known as a spectrogram~\cite{rs2010}.
The spectrogram has been used for speech forensics using a transformer neural network~\cite{vaswani_2017} or a \gls{cnn}~\cite{spec_vgg_sincnet_28, amit_ei_paper, spec_cqcc_resnet_se_27,spec_cnn_telefor,acm_21_logspec}.

These methods for synthetic speech detection lack interpretability~\cite{explainable_ai_2022,cvprw_2022}, \ie~they do not provide any reasoning behind how a detector produces its output. 
Some work has been done in providing explanations for synthetic speech detectors output~\cite{exp_cqcs_slt20, exp_cnn_slt_2018,exp_fake_22, exp_fake_det_icassp22, exp_halpern20_interspeech}.
For example, Ge~\etal~\cite{exp_fake_det_icassp22} analyzed the detector behaviour using Shapley Additive Explanations~\cite{shap_nips} and Chettri~\etal~\cite{exp_cnn_slt_2018} used Local Interpretable Model-Agnostic Explanations~\cite{lime_kdd} to provide artifacts that contribute most to synthetic speech detectors output.
Tak~\etal~ in ~\cite{exp_cqcs_slt20} analyzed different sub-band components of the spectrum of a speech signal obtained using Fourier transform to explain which sub-band contributes most to synthetic speech artifacts.
These methods mainly focus on post analysis of detectors to add explanation behind its working.



In this paper, we examine spectrogram representations of speech to obtain disentangled representations using a Variational Auto Encoder network.
These disentangled representations when visualized provide explanation behind working of detector.

\subsection{Disentangled Representation Learning}
Disentangled representation learning methods leverage the idea that it is possible to divide learned representations into multiple explainable components~\cite{disentagled_representation,Zhu_2021_CVPR}. 
This concept has been used in generative networks for image generation to control regions of synthetic images~\cite{ icpr_2021_image_generation, cvpr_20_mixnMatch}. 
Some methods have also used disentangled representation learning for voice conversion, and voice style transfer during speech synthesis~\cite{icassp_voice_style_transfer, acm_sigsac_2020,iterspeech_voice_conversion, interspeech_22_anonymization}.

The use of disentangled representation learning in forensic applications is mostly limited to face anti-spoofing systems~\cite{eccv_disentagled_2020,eccv_disentagled_2020_generic}. 
Zhang~\etal~proposed a one stage disentanglement network using autoencoder for face anti-spoofing~\cite{eccv_disentagled_2020}.
Wang~\etal~also used an autoencoder neural network but with a two stage training strategy to generalize performance to unseen face spoofing attacks~\cite{wacv_disentangled_2022}.
Two stage disentanglement autoencoder proposed in ~\cite{wacv_disentangled_2022} showed better generalization to detect unknown face spoofing attacks.
We propose a variational autoencoder architecture and use two stage training to learn disentangled representation for speech forensics.



\section{Proposed Method}\label{sec:method}

In this paper we propose ~\methodfullname~ to detect if a speech signal under analysis is bona fide or synthetically generated.
We do so by exploiting variational autoencoders that provide disentangled representation of the input speech signal.
We refer to these representations as disentangled because they depend on only one characteristic of speech signal \ie~whether speech signal is bona fide or synthetic.
Hence, these representations disentangle when visualized for bona fide and synthetic speech adding explanation behind the working of detector.
The disentangled representation is also used to compute an activation map that highlights which regions of the input are used for classification.
Together with disentangled representation and an activation map \methodname~ sheds light on the working principle of the detector, making it more explainable.

In this section, we first provide an overview of the overall proposed method.
Then we report additional details about the training procedure.

\subsection{Overall Approach}\label{sec:pipeline}
An autoencoder maps an input data point to a representation vector.
A variational autoencoder maps input to a distribution from which the data point could have been generated~\cite{vae_first_paper_2013}. Wang~\etal and Zhang~\etal used autoencoder network for face anti-spoofing~\cite{wacv_disentangled_2022, eccv_disentagled_2020}.
We modified their autoencoder architecture, added stochastic variational inference and learning ~\cite{vae_first_paper_2013}, and optimized the obtained variational autoencoder for processing spectrograms. 
Contrary to face images, spectrograms are 2-D single channel inputs and require different number of layers, kernel size, and dimension of representation vector. The architecture of our optimized variational autoencoder can be found in our code.

\begin{figure*}[!t]
    \centering
    \includesvg[width=1.0\linewidth]{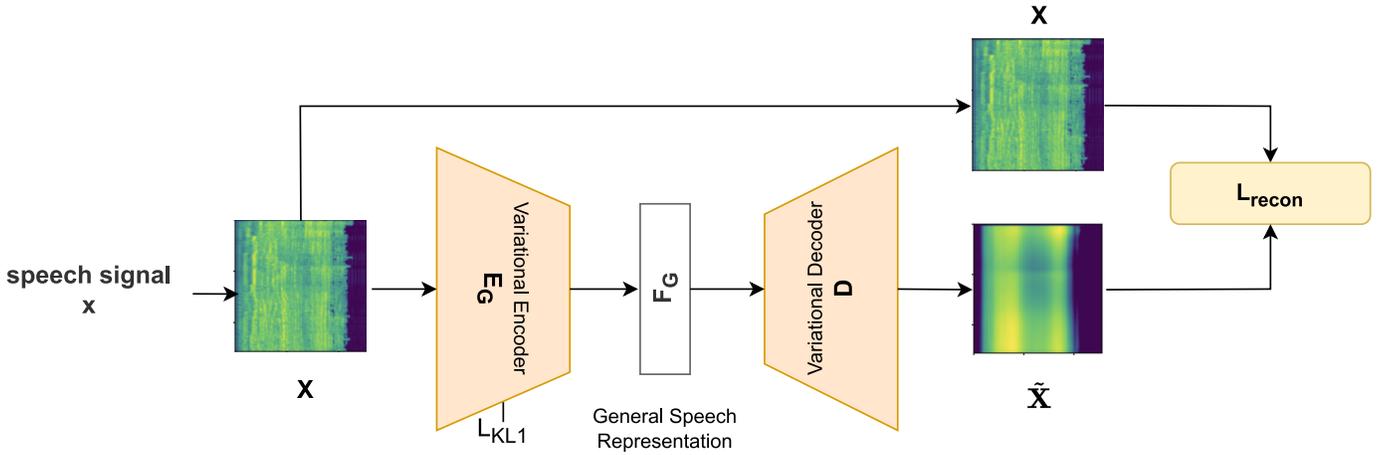}
    \caption{Block diagram of ~\methodname~ for Training Stage 1: The General Representation Network.}
    \label{fig:stage1}
\end{figure*}
\begin{figure*}[!t]
    \centering
    \includesvg[width=1.0\linewidth]{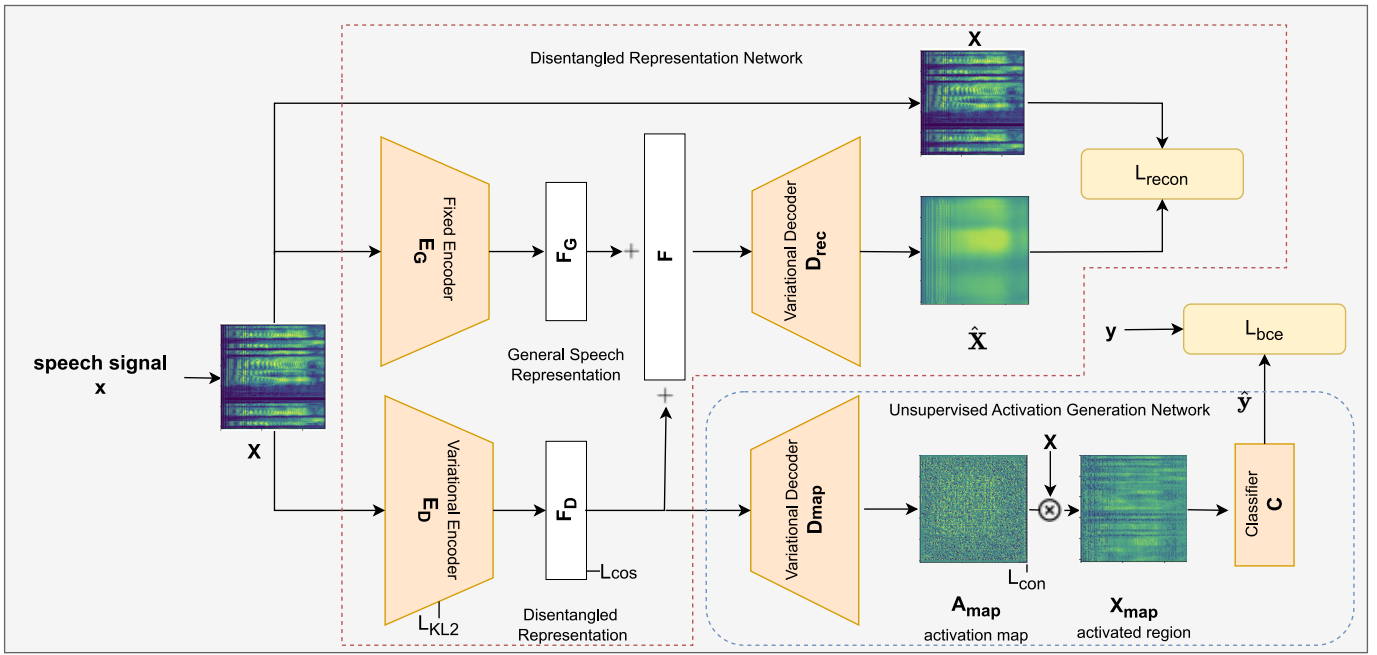}
    \caption{Block diagram of ~\methodname~ for Training Stage 2: The Disentangled Representation Network and The Unsupervised Activation Generation Network.}
    \label{fig:stage2}
\end{figure*}
\begin{figure*}[!ht]
    \centering
    \includesvg[width=1.0\linewidth]{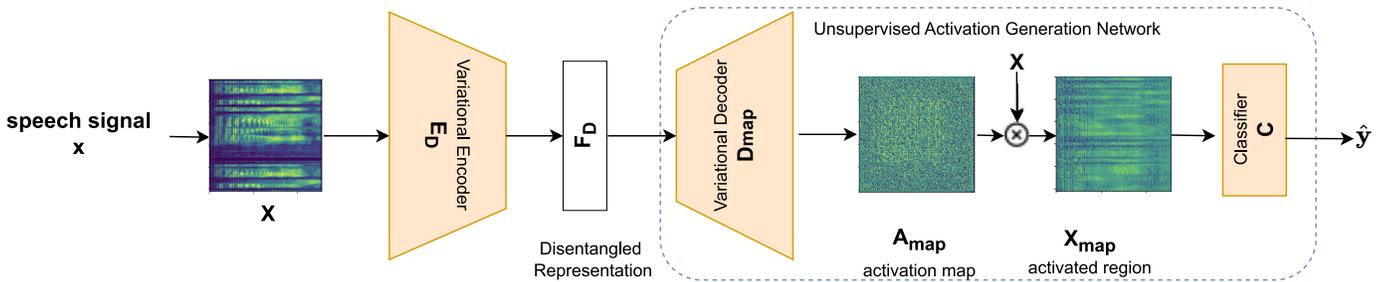}
    \caption{Block diagram of ~\methodname~ for evaluation/inference.}
    \label{fig:inference}
\end{figure*}

\color{black}
Given a time domain speech signal $\Vec{x}$, we compute the magnitude of its Short Time Fourier Transform (STFT) using a Hanning window of size 25 ms with a shift of 10 ms to obtain the spectrogram~\cite{rs2010}.
We used these parameters following previous methods which use spectrogram for general audio classification and speech forensics tasks~\cite{koutini_2021, msm_mae_niizumi}.
We convert the frequency axis of the spectrogram from the Hertz scale
to the mel-scale~\cite{mel}.
The mel-scale correlates better with the human auditory system as compared to the Hertz scale~\cite{mel}.
The conversion between the Hertz frequency scale $f_\text{Hz}$ and the mel frequency scale $f_\text{mel}$~\cite{mel} is obtained by
\begin{align}
    f_\text{mel} = 2595 \cdot \log_{10}{\left(1 + \frac{f_\text{Hz}}{700}\right)}
\end{align}
We represent the mel-scale spectrogram by  $\Vec{X}$.
Our claim is that the mel-spectrogram $\Vec{X}$ can be decomposed into two parts that contain complementary pieces of information about the speech signal. 
The first type of information is  general speech information which is common in both bona fide and synthetic speech signals.
We denote this by $\Vec{F}_\text{G}$ (see ~\cref{fig:stage1} and ~\cref{fig:stage2}) and refer to it as the general speech representation.
The second type of information discriminates bona fide speech from synthetic speech.
We denote it by $\Vec{F}_\text{D}$ (see ~\cref{fig:stage2}) and refer to it as the disentangled representation.
The disentangled representation discriminates w.r.t whether the speech signal is bona fide or synthetic.

We use a two stage training approach to obtain $\Vec{F}_\text{G}$ and $\Vec{F}_\text{D}$.
In  Training Stage 1, we use the General Representation Network described in ~\cref{sec:stage1} (\cref{fig:stage1}) to 
obtain $\Vec{F}_\text{G}$. 
Once we obtain $\Vec{F}_\text{G}$, we fix the weights of the encoder $E_{G}$ that provides general speech representation \ie 
 ~$\Vec{F}_\text{G}$ in ~\cref{fig:stage2} and then train the Disentangled Representation Network and the Unsupervised Activation 
Generation Network described in ~\cref{sec:stage2} (\cref{fig:stage2}) to obtain $\Vec{F}_\text{D}$ and an activation map $\Vec{A}_\text{map}$.
$\Vec{A}_\text{map}$ shows regions of spectrogram that contain discriminative features for detecting synthetic speech.

During evaluation/inference, we do not need to obtain the general speech representation \ie~ $\Vec{F}_\text{G}$.
We need the disentangled representation \ie~ $\Vec{F}_\text{D}$ to decide whether a speech signal is synthetic or bona fide.
The block diagram for our evaluation/inference stage is shown in ~\cref{fig:inference}.

\subsection{Training Stage 1}\label{sec:stage1}
In the first training stage, we train the General Representation  Network shown in \cref{fig:stage1} for the self-supervised task of reconstructing the input spectrogram that contains the common speech information~\cite{vae_first_paper_2013,wacv_disentangled_2022}. 
 We do not use any synthetic speech signals during this stage.
 We train only on real speech signals. 
 This ensures that the method does not learn any discriminative features during this stage. 

The variational autoencoder consists of an encoder $E_{G}$ and a decoder $D$.
The Encoder $E_{G}$ produces a general speech feature vector $\Vec{F}_\text{G}$ from the spectrogram $\Vec{X}$. 
The Decoder \textbf{$D$} uses $\Vec{F}_\text{G}$ to reconstruct the spectrogram of the common speech information.
We denote the reconstructed spectrogram by $\Tilde{\Vec{X}}$, where $\Tilde{\Vec{X}}= D(E_{G}(\Vec{X}))$.

We train the network using a weighted average of two loss functions: the reconstruction loss; the \gls{kl}-divergence loss.
The reconstruction loss $\mathcal{L}_{recon}$ is defined as
\begin{align}
\mathcal{L}_{recon} = 
E [|| \Vec{X}- \Tilde{\Vec{X}} ||_{2}^{2}]
,\end{align}
where $E$ is the expected value/mean of the square error 
between the input $\Vec{X}$ and the output $\Tilde{\Vec{X}}$.
The \gls{kl}-divergence loss $\mathcal{L}_{KL1}$ is the typical loss function used for variational autoencoders and is defined as
\begin{align}\mathcal{L}_{KL1} = \Sigma_{i} KL (q_{i}^{g}(z^{g}|x) || p(z^{g}|x)) \end{align}
Here, $q^{g}$ is the distribution learned by encoder $E_{G}$, $p$ is the true prior distribution, $x$ is an instance of input spectrogram $\Vec{X}$, and $z^{g}$ is an instance \ie~ $\Vec{F}_\text{G}$ sampled from the learned distribution.
We followed reparameterization mentioned in ~\cite{vae_first_paper_2013} for variational autoencoder $E_{G}$ and $\mathcal{L}_{KL1}$.

\subsection{Training Stage 2}\label{sec:stage2}
In the second training stage, we train the Disentangled Representation Network and the Unsupervised Activation Generation Network shown in ~\cref{fig:stage2} to obtain the disentangled representation and the activation map.
The Disentangled Representation Network processes the spectrogram $\Vec{X}$ using two identical variational autoencoders $E_{G}$ and $E_{D}$ to obtain a general speech representation $\Vec{F}_\text{G}$ and 
a disentangled representation $\Vec{F}_\text{D}$.
The goal of the second training stage is to learn interpretable features that differentiate/disentangle bona fide speech from synthetic speech.

We fix the weights of $E_{G}$ to the weights obtained from Training Stage 1 (\cref{sec:stage1}).
Fixing the weights of $E_{G}$ fixes the network used for obtaining the general speech representation from the input speech signal.
Note that we hypothesize that both bona fide and synthetic speech have these general features and they are common for both classes.
We concatenate the general feature vector $\Vec{F}_\text{G}$ and the disentangled feature vector $\Vec{F}_\text{D}$ to form $\Vec{F}$.
We process $\Vec{F}$ using a decoder $D_{rec}$ to reconstruct the spectrogram. 
We denote the reconstructed spectrogram as $\hat{\Vec{X}}$ (see \cref{fig:stage2}).
Note that while learning the general speech representation in Training Stage 1 we used a different decoder $D$ (see \cref{fig:stage1}). 

To train this network we minimize a weighted average of multiple loss terms.
To generate spectrograms similar to the original input spectrograms, we use the reconstruction loss $\mathcal{L}_{recon}$ described in ~\cref{sec:stage1}.
We also use the \gls{kl}-divergence loss $\mathcal{L}_{KL2}$ for the encoder \textbf{$E_{D}$} defined as
\begin{align}\mathcal{L}_{KL2} = \Sigma_{i} KL (q_{i}^{d}(z^{d}|x) || p(z^{d}|x)) \end{align}
Here, $q^{d}$ is the distribution learned by encoder $E_{D}$, $p$ is the true prior distribution, $x$ is an instance of input spectrogram $\Vec{X}$, and $z^{d}$ is an instance \ie~ $\Vec{F}_\text{D}$ sampled from the learned distribution.
To disentangle $\Vec{F}_\text{D}$ for bona fide and synthetic speech signal we use the CosFace loss $\mathcal{L}_{cos}$ proposed in~\cite{cosface_loss_cvpr}.
This loss minimizes intra-class variance of $\Vec{F}_\text{D}$ and maximizes inter-class variance by maximizing the decision margin in the angular space~\cite{cosface_loss_cvpr}.

The Unsupervised Activation Generation Network further processes the disentangled feature $\Vec{F}_\text{D}$ to generate an activation map that highlights the spectrogram region that is captured by the disentangled representation.
We denote the activation map by $\Vec{A}_\text{map}$.
\methodname~ processes $\Vec{F}_\text{D}$ using the Decoder $D_{map}$ to create the activation map $\Vec{A}_\text{map}$. 
Following ~\cite{wacv_disentangled_2022}, to avoid trivial identity activation map, we impose $L_{1}$ regularization on the activation map for bona fide speech signal.
We refer this loss as $\mathcal{L}_{con}$ and it is defined as
\begin{align}\mathcal{L}_{con} = E[||\Vec{A}_\text{map}||_{1}]\end{align}
where $E$ is the expected value from all bona fide speech signals.
This loss helps to ensure that the activation map is not an identity map activating all regions of spectrogram.
We take dot product of the activation map $\Vec{A}_\text{map}$ and the spectrogram $\Vec{X}$ to obtain the activated spectrogram map 
$\Vec{X}_\text{map}$.
Following previous work in face anti-spoofing~\cite{wacv_disentangled_2022,eccv_disentagled_2020}, we process $\Vec{X}_\text{map}$ using a classifier $C$ to obtain probability of the speech signal being synthetic which is used to determine the prediction label \ie, \textbf{$\hat{y}$}.
To train the classifier $C$, we use the binary cross entropy loss $\mathcal{L}_{bce}$ defined as
\begin{align}\mathcal{L}_{bce} = E[ -y\cdot\log(\hat{y})+(1-y)\cdot\log(1- \hat{y})]\end{align}
where $E$ is expected value from all speech signal, $\hat{y}$ is the prediction label from classifier $C$ and $y$ is ground truth label \ie equal to 1 for synthetic speech and 0 for bona fide speech.


\section{Experiments}\label{sec:setup}

In this section, we describe the dataset used for our experiments, the objective metrics, and the parameters used in our implementation.

\subsection{Dataset}
We use the Logical Access (LA) part of the ASVspoof2019 dataset \cite{asvspoof19, asvdata_2019}.
This dataset contains approximately 121.5k speech signals. 
These signals are divided into a training set $D_{tr}$, a validation set $D_{dev}$, and an evaluation set $D_{eval}$~\cite{asvspoof19} in the approximate ratio of 1:1:3.
Each set is highly unbalanced w.r.t the number of bona fide and synthetic speech signals.
There are approximately 89\% synthetic speech signals and 11\% bona fide speech signals in each set.
These speech signals are encoded using the \gls{flac} format~\cite{flac}.

We select this dataset for evaluation because it contains
synthetic speech signals from unknown synthesizers in the evaluation set $D_{eval}$.
Investigating this dataset helps us to identify if our disentangled representation generalizes to 
unknown synthesizers.
The dataset contains synthetic speech signals from 19 speech synthesizers A01 to A19.
Two pairs of synthesizers namely (A04, A16), and (A06, A19) have the same underlying architectures but are trained on different datasets.
Therefore, overall there are only 17 different speech synthesizers.
All 17 speech synthesizers can be categorized into one of the three categories based on the type of generation: neural networks, vocoders, and waveform concatenation \cite{asvdata_2019}.
There are 63.9k synthetic speech signals in the $D_{eval}$ set, out of which 61.5k synthetic speech signals are generated from 11 unknown synthesizers A07 to A18 (except A16) that are not present in $D_{tr}$ or $D_{dev}$.
The training and validation sets contain synthetic speech signals from methods A01 to A06. 
The bona fide speech signals in the training, validation, and testing sets are recorded from human speakers which do not overlap among the three sets.

\subsection{Evaluation Metrics}
We use \gls{eer} as our performance metric for evaluation. It is the recommended metric for comparison in the ASVspoof2019 Challenge~\cite{asvspoof19,asvdata_2019}.
We obtain \gls{eer} from \gls{roc} by finding the rate where \gls{fnr} and \gls{fpr} are equal.
Lower the \gls{eer}, the better is the performance of a method.
We also report detection accuracy \cite{tharwat2021} which is percentage of total correct classifications out of all the classifications made.



\subsection{Implementation Details}\label{sec:train-and-test}
For Training Stage 1 (\ie~ to find the general speech representation) we use speech signals from the Audioset dataset~\cite{audioset}.
We fixed the dimension of general speech representation vector (\ie~ $\Vec{F}_\text{G}$) to 512 dimensions.
We use the Adam optimizer~\cite{adam} with an initial learning rate of $10^{-3}$, a decay rate of $5\times 10^{-7}$ and trained the network using a batch size of 256.
We trained the network for $\approx$ 333K iterations until convergence (the loss is
not changing and is below 0.37).
We use $\mathcal{L}_{rec}$ and \gls{kl}-divergence loss $\mathcal{L}_{KL1}$ described in ~\cref{sec:stage1} for training the variational encoder $E_{G}$.
Therefore, the loss for Training Stage 1 ($\mathcal{L}_{stage1}$) is given by
\begin{align}\mathcal{L}_{stage1} = \mathcal{L}_{recon}+
\mathcal{L}_{KL1} 
\end{align}

For the second stage of training we use the AdamW~\cite{adam_w} optimizer with an initial learning rate of $10^{-4}$ and a weight decay of $10^{-3}$ for training.
We fixed the dimension of disentangled speech representation vector (\ie~ $\Vec{F}_\text{D}$) to 512 dimensions.
We fixed the weights of the variational encoder $E_{G}$ and trained using the sum of five losses- $\mathcal{L}_{bce}$, $\mathcal{L}_{cos}$, $\mathcal{L}_{con}$, $\mathcal{L}_{KL2}$, and $\mathcal{L}_{recon}$. 
These losses are defined in ~\cref{sec:stage1} and ~\cref{sec:stage2}. 
Therefore the loss for Training Stage 2 ($L_{stage2}$) is  
\begin{align}\mathcal{L}_{stage2} = 
\mathcal{L}_{recon}+
\mathcal{L}_{cos} + 
\mathcal{L}_{con} +
\mathcal{L}_{KL2} + 
\mathcal{L}_{bce}  \end{align}
We trained the second stage for 100 epochs with a batch size of 100 on the training set $D_{tr}$ from ASVspoof2019~\cite{asvspoof19}.
We select the model with the best balanced accuracy on validation set $D_{dev}$ for evaluation.


\begin{figure*}[!t]
    \centering
    \includegraphics[width=1.0\linewidth]{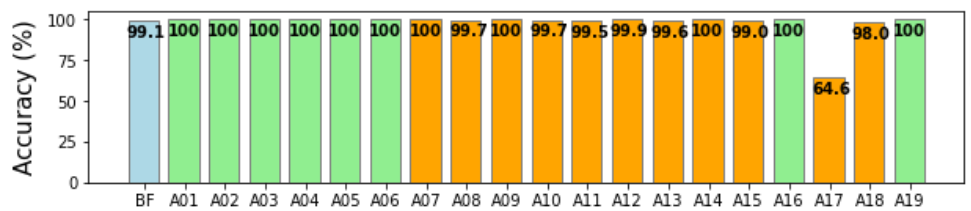}
    \caption{Detection accuracy of \methodname~ on bona fide speech signal (blue), and synthetic speech from known (green) and unknown (orange) synthesizers in the 
    of the ASVspoof2019 dataset. A16 and A19 have same architecture as A04 and A06.}
    \label{fig:ind-accuracy}
\end{figure*}
\begin{figure*}[!t]
    \centering
    \includegraphics[width=1\linewidth]{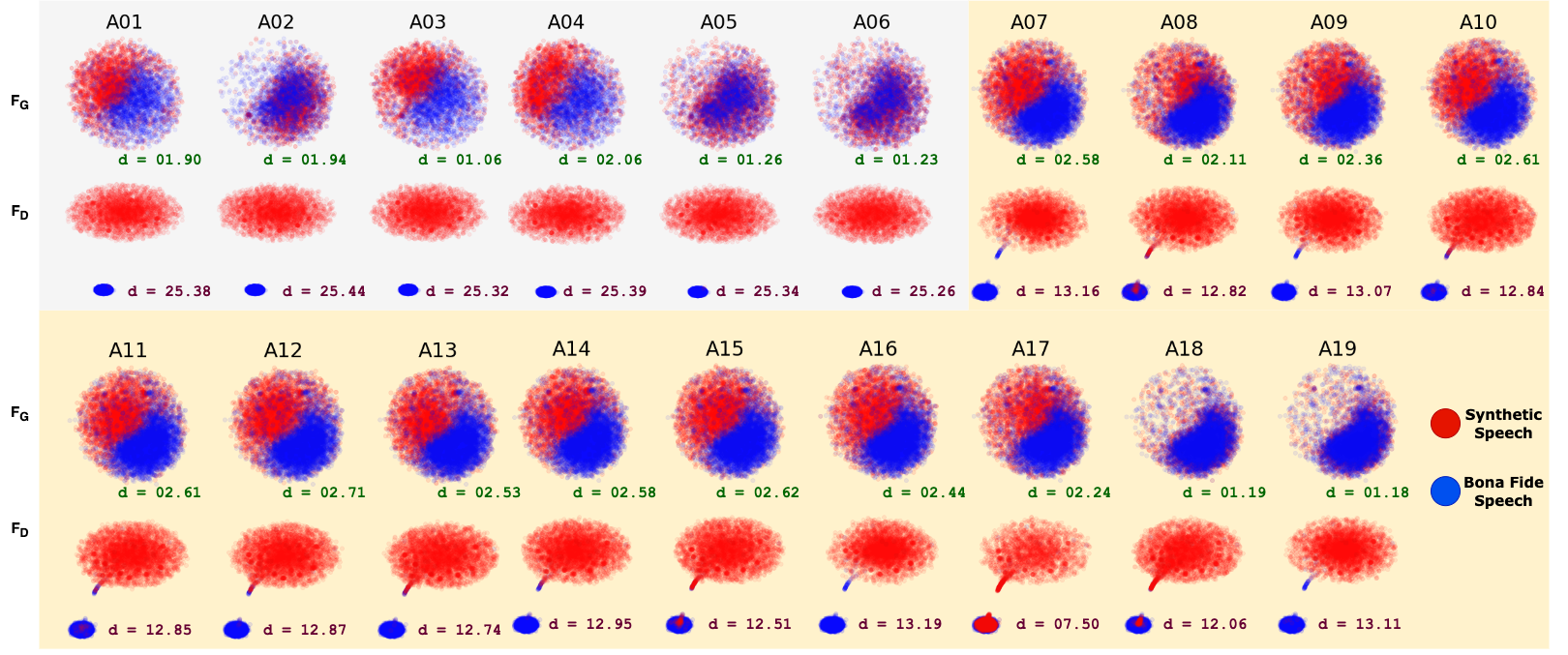}
    \caption{2-D t-SNE visualization of general speech representation ($\Vec{F}_\text{G}$) and disentangled representation ($\Vec{F}_\text{D}$) obtained by \methodname~ for A01 to A06 speech synthesizers in validation set and A07 to A19 speech synthesizers in the evaluation set of the ASVspoof2019 dataset.
    `\textbf{d}' is the 
    Mahalanobis distance between the distribution of bona fide speech representation and 
    synthetic speech representation from a synthesizer. `\textbf{d}' in green is obtained from $\Vec{F}_\text{G}$ and `\textbf{d}' in brown is obtained from $\Vec{F}_\text{D}$. }
    \label{fig:visual}
\end{figure*}
\section{Experimental Results}\label{sec:results}
This section describes the results of our experiments.

\subsection{Synthetic Speech Detection}
In this experiment, we evaluated the ability of \methodfullname~ to distinguish between bona fide and synthetic speech. 
We take synthetic speech signals from all the synthesizers in $D_{dev}$ and $D_{eval}$ set of the ASVspoof2019 dataset \cite{asvspoof19}.
The synthetic speech signals are generated from 19 synthesizers (from A01 to A19), out of which synthetic speech signals from 6 synthesizers (from A01 to A06) were present in the training set.
Synthesizer A16 and synthesizer A19 have same underlying architecture as synthesizer A04 and A06, respectively.
Therefore there are 17 different synthesizers, 6 known and 11 unknown synthesizers.
For bona fide class we used the union of bona fide speech signals from $D_{dev}$ and $D_{eval}$ set. 
These are from 58 different human speakers that do not overlap with the speakers from the training set.

\cref{fig:ind-accuracy} shows the detection accuracy of ~\methodname~ on speech  belonging to each one of the synthesizers as well as bona fide speech.
In total there are approximately 10k bona fide speech signals and about 86k synthetic speech signals. 
We observe from ~\cref{fig:ind-accuracy} that \methodname~ has perfect detection accuracy for all known and unknown synthesizers and bona fide speech except for the unknown synthesizer A17. 
Synthesizer A17 
belongs to the most challenging class according to the ASVspoof2019 Challenge result~\cite{asvspoof_2021}.
A significant portion in the starting of the signal is non speech segment (silence).
For example, the silence in the start covers on average around 26\% duration of the total duration for each signal from synthesizer $A17$.
Speech signal from other unknown synthesizers such as synthesizer $A14$ have on average same total duration but the the silence in the start covers on average around 8\% of the total duration.
The accuracy reported in \cref{fig:ind-accuracy} for synthesizer $A17$ corresponds to the one obtained after removing silence segment from the start of the signal generated from $A17$. 
With the silence segment the detection accuracy for synthetic speech signal from synthesizer $A17$ was less than 10\%.
Therefore, removing silence segment from starting improved performance of \methodname~ on $A17$ class.
However, it is still less comparative to other classes.
A possible reason could be after removing non-speech (silent) segment from starting, \methodname~ had on average duration of only 2.5 seconds. 
Hence, there is less speech segment to analyze and that might be the reason for lower performance of \methodname~ on class $A17$.

\subsection{Visualizing General and Disentangled Features}
In this experiment, we visualized the general speech representations $\Vec{F}_\text{G}$ and 
disentangled representations $\Vec{F}_\text{D}$ learnt by \methodname. 
$\Vec{F}_\text{G}$ represents those features that are common to both bona fide and synthetic speech, hence during visualization we expect overlapping of $\Vec{F}_\text{G}$ for synthetic and bona fide speech signals.
$\Vec{F}_\text{D}$ represents those features that differentiate synthetic speech from bona fide speech, hence during visualization we expect either very minimal or no overlapping of $\Vec{F}_\text{D}$ for synthetic and bona fide speech signals.

For 2D visualization of the feature vectors $\Vec{F}_\text{G}$ and $\Vec{F}_\text{D}$, we first projected them from 512-D to 2-D using T-distributed Stochastic Neighbor Embedding (t-SNE) \cite{maaten2008}.
\cref{fig:visual} shows a visualization for both $\Vec{F}_\text{G}$ and $\Vec{F}_\text{D}$ for each of the synthesizers present in the ASVspoof2019 dataset.
The red color corresponds to synthetic speech while the blue color corresponds to bona fide speech.
\begin{figure}[!b]
    \centering    
    \includesvg[width=0.4\linewidth]{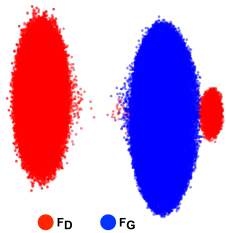}
    \caption{2-D t-SNE visualization showing general speech representation learnt by \methodname~ is different from disentangled representation.}
    \label{fig:visualfgfd}
\end{figure}
In ~\cref{fig:visual}, we can observe that the general speech representations overlap and the disentangled representations do not overlap except for very few speech samples.
This is true for all the known synthesizers and 10 out of 11 unknown synthesizers.
The disentanglement can also be shown quantitatively. 
We compute the Mahalanobis distance \cite{mahalanobis1936generalized} (denoted by \textbf{d}) between the distribution of bona fide and synthetic speech representation for each of the synthesizer.
Larger Mahalanobis distance indicates the two distributions are more dissimilar.
\cref{fig:visual} shows that the Mahalanobis distance obtained from $\Vec{F}_\text{D}$ (in brown) is very large compared to Mahalanobis distance obtained from $\Vec{F}_\text{G}$ (in green).
Therefore, feature $\Vec{F}_\text{D}$ disentangle synthetic speech from bona fide speech.
For synthesizer A17, 
visualization of $\Vec{F}_\text{G}$ representation is as expected and \textbf{d} obtained from $\Vec{F}_\text{G}$ is also low.
But visualization of $\Vec{F}_\text{D}$ representation shows that many of the synthetic speech signals are not disentangling from bona fide speech signal.
Notice that 
disentangled representation $\Vec{F}_\text{D}$ for several synthetic speech signals from synthesizer $A17$ are on top of the cluster created by disentangled representation $\Vec{F}_\text{D}$ of bona fide speech.
This is also reflected by low detection accuracy on synthesizer A17 in ~\cref{fig:ind-accuracy} and lower value of \textbf{d} obtained from $\Vec{F}_\text{D}$ for synthesizer A17.
Therefore, low detection accuracy for any unknown synthesizer can directly be attributed to overlapping in disentangled representation $\Vec{F}_\text{D}$ and lower value of \textbf{d} obtained from $\Vec{F}_\text{D}$.
We observe that visualizing disentangled features adds reasoning to the decision made by ~\methodname.

Some of the previous methods proposed for disentanglement representation learning have argued that it is important to verify that feature vectors $\Vec{F}_\text{G}$ and $\Vec{F}_\text{D}$ do not overlap and are different to ensure robust disentangled representation learning~\cite{wacv_disentangled_2022, eccv_disentagled_2020_generic}.
This ensures that learned disentangled representation are different from general speech representation.
In ~\cref{fig:visualfgfd}, we show  that feature vectors $\Vec{F}_\text{G}$ and $\Vec{F}_\text{D}$ do not overlap and are different. 
\begin{figure*}[!t]
    \centering    \includegraphics[width=0.9\linewidth]{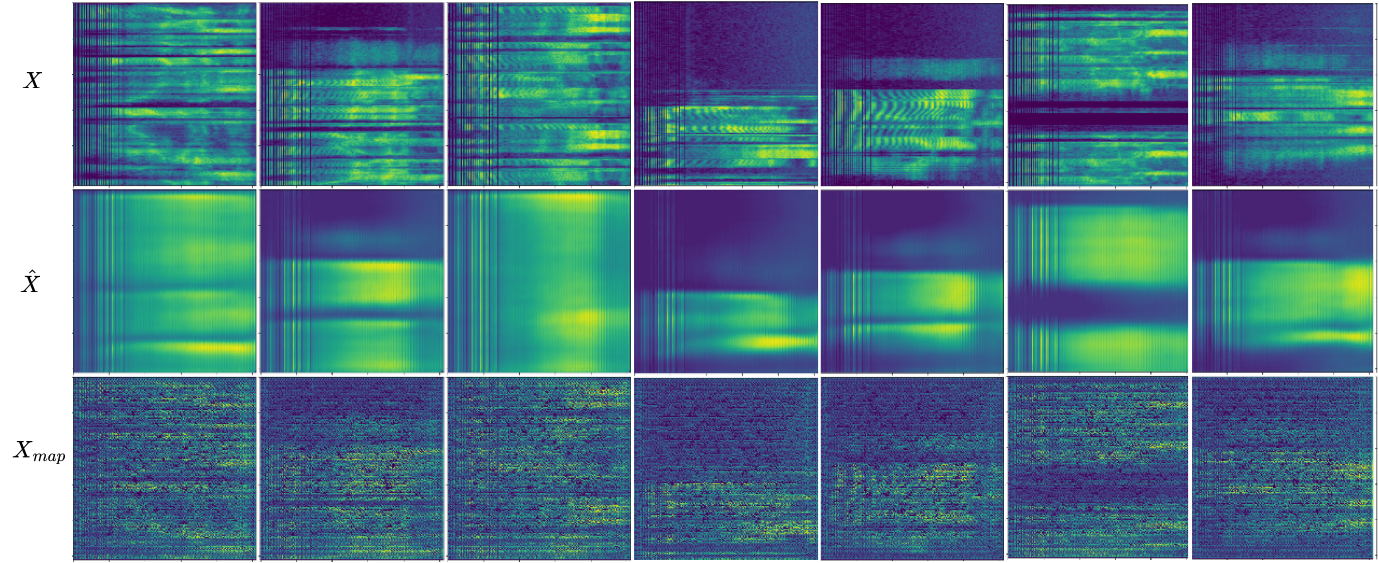}
    \caption{Visualization of input spectrogram:$\Vec{X}$, reconstructed spectrogram:$\hat{\Vec{X}}$, and activated region: $\Vec{X}_\text{map}$.}
    \label{fig:spect}
\end{figure*}
\begin{table}[!t]
    \centering
    \small
    \caption{\gls{eer}\% of \methodname~ and 16 other methods 
    for synthetic speech detection using the ASVspoof2019 dataset.}

\begin{tabular}{@{\extracolsep{-4pt}}lcccc}
    \toprule
     Method Name & Feature & Network & EER \\\midrule
     $B01$ & CQCC  & GMM & 8.09\% \\
     $B02$ & LFCC   & GMM & 9.57\% \\
     $S01$ & Spectrogram & VGG & 10.52\%\\
     $S02$ & Log-Spectrogram & MesoInception & 10.02\%\\
     $S03$ & Spectrogram & CNN & 9.57\%\\
     $S04$ & Spectrogram+CQT & VGG+SincNet & 8.01\% \\
     $S05$ & Spectrogram$_{CQOST}$ & DNN & 8.04\% \\
     $S06$ & Spectrogram$_{ICQCC}$ & DNN & 7.70\% \\
     $S07$ & Spectrogram$_{CQT}$ & Transformer & 7.50\% \\
     $S08$ & Spectrogram$_{CQT}$ & MesoNet & 7.42\% \\
     $S09$ & Spectrogram$_{CQT}$ & LSTM &  7.16\%\\
     $S10$ & Spectrogram$_{CQT}$ & LCNN-Attention & 6.76\% \\
     $S11$ & Spectrogram$_{CQT}$ & ResNet18 & 6.55\%\\
     $S12$ & Spectrogram$_{CQT}$ & LCNN & 6.35\% \\
     $S13$ & Spectrogram$_{CQT}$ & LCNN+LSTM & 6.23\% \\
     $S14$ & Mel-Spectrogram & PaSST & 5.26\% \\
     \textbf{\methodname} & Mel-Spectrogram & VAE & \textbf{2.16\%} \\
     \bottomrule
     \end{tabular}
    \label{tab:comparison}
\end{table}
\subsection{Comparison with Existing Methods}
In this experiment, we compare the performance of our method with 16 other methods for synthetic speech detection.
The proposed \methodfullname~ uses 
spectrogram to detect synthetic speech. 
Therefore, for comparison, we only include methods that either use spectrogram or features obtained from spectrogram or are proposed in ASVspoof2019 Challenge as baselines~\cite{asvspoof19}.
The methods proposed in the ASVspoof2019 Challenge as baselines~\cite{asvspoof19} are methods 
$B01$ and $B02$ \cite{asvspoof19}.
Both of these methods use \gls{gmm} for processing hand-crafted features such as \gls{lfccs}, and \gls{cqccs}.
The remaining 14 methods ($S01$ to $S14$) that we use for comparison either process spectrogram or features obtained from spectrogram using different neural networks.

We can obtain spectrogram by processing time domain signal using either \gls{stft} or using \gls{cqt}.
In \cref{tab:comparison}, Spectrogram denotes the one obtained using \gls{stft} and Spectrogram$_{CQT}$ denotes the one obtained using \gls{cqt}.
$S05$ uses Spectrogram$_{CQOST}$ and $S06$ uses Spectrogram$_{ICQCC}$ \cite{tifs_cqost_da}.
These features are obtained by using octave sub-banding and applying \gls{dct} to either Spectrogram$_{CQT}$ or inverted Spectrogram$_{CQT}$.
$S02$ \cite{in_the_wild} and $S14$ \cite{bartusiak_theasis} map the spectrogram obtained using \gls{stft} to logarithmic and mel scale, respectively.
$S01$ \cite{spec_vgg_sincnet_28} uses VGG and $S02$ \cite{in_the_wild} uses Inception neural network to process features \cite{vgg_first_paper}.
Method $S03$ \cite{spec_cnn_telefor} process Spectrogram using a \gls{cnn}.
$S04$ \cite{spec_vgg_sincnet_28} uses fusion of two features, namely, Spectrogram and \gls{cqt} \cite{rs2010} to detect synthetic speech (refer as Spectrogram$+CQT$ in \cref{tab:comparison}).
$S05$ and $S06$ \cite{tifs_cqost_da} use \gls{dnn}  to process Spectrogram$_{CQOST}$ and Spectrogram$_{ICQCC}$ feature, respectively.
There are several version of method $S02$, and $S07$ to $S13$. Each version is trained on a different feature derived from spectrogram (\eg~ Spectrogram$_{CQT}$ and Log-Spectrogram). 
In our comparison, we report the \gls{eer} of the best performing version.
More details about methods $S02$, $S07$ to $S13$ and their different versions can be found in \cite{in_the_wild}.
Bartusiak~\etal~ proposed several transformer neural networks to process Mel-Spectrogram and detect synthetic speech in \cite{bartusiak_theasis}.
Method $S14$ is the best individual performing transformer neural network 
in \cite{bartusiak_theasis}.
Similar to \methodname~, $S14$ also processes Mel-Spectrogram.
$S14$ is first trained using self-supervision on a large audio dataset \ie Audioset \cite{audioset}.
For inferencing, it uses 85.3M parameters, almost twice the number of parameters used by \methodname.

~\cref{tab:comparison} summarizes our results.
We can notice that $S14$ with \gls{passt} \cite{koutini_2021} has better performance among all the comparison methods.
\methodname~ has better performance than all the 16 methods.
\methodname~ provides an improvement of more than 3 percentage points in \gls{eer} from $S14$ which has twice the number of parameters.
There is around 6 percentage points and 7 percentage points improvement in \gls{eer} from baseline methods $B01$ and $B02$, respectively.
Overall the working principle of \methodname~ is more understandable by visualizing the disentangled features and it has a significant improvement in \gls{eer} than several existing methods such as $S04$.

\subsection{Performance in practical scenarios}
First, we investigate performance of \methodname~ when synthetic speech is uploaded on social platforms. 
Different social platforms use different compression standards.
For this evaluation we use around 534K speech signals from the evaluation part of the ASVspoof2021 DeepFake (DF) dataset \cite{asvspoof_2021}.
It contains speech signals from the evaluation set of ASVspoof2019 dataset \cite{asvspoof19} and 2018 and 2020 Voice Conversion Challenge (VCC) datasets \cite{vcc_2018, vcc_2020}.
Speech signals consist of uncompressed, single compressed and double compressed speech signals.
Single compression is done using MP3, AAC, and OGG compression standards \cite{mpeg4_book, isomp3, isoaac} at both high and low variable bit rate.
Double compression is done from two standards MP3 and OGG at low variable bit rate to AAC standard at high variable bit rate.
Yadav~\etal showed that detectors with EER less than 5\% on uncompressed speech have almost random accuracy (EER 50\%) when tested on compressed speech~\cite{assd_2023}. 
\methodname~ has detection accuracy of 92.25\% with an \gls{eer} of 32.28\% on the evaluated dataset with compressed speech and hence \methodname~ can detect synthetic speech uploaded to different social platforms.

Some existing methods trained on ASVspoof2019 have been shown to use silence region for detecting synthetic speech \cite{silence_gold}. A simple attack such as removing silence can lead to failure and random performance of such detectors.
Therefore, secondly, we investigate the performance of \methodname~ when non speech elements \ie~ silence regions are removed from each speech signal.
For this we use the hidden part of the ASVspoof2021 DeepFake (DF) dataset \cite{asvspoof_2021}. It consists of 18.5K speech signal randomly selected from evaluation set of ASVspoof2019 \cite{asvdata_2019} and 2018 and 2020 Voice Conversion Challenge (VCC) datasets \cite{vcc_2018, vcc_2020}.
\methodname~ has detection accuracy of 84.09\% on hidden part with an \gls{eer} of 38.28\%. Hence, \methodname~ is robust to simple attacks such as removing silence.

\subsection{Reconstructed Spectrograms and Attention}
We randomly select 7 speech signals from the evaluation set of the ASVspoof2019 dataset.
In ~\cref{fig:spect}, we show their spectrograms $\Vec{X}$, the spectrograms reconstructed by $D_{rec}$ using feature vector $\Vec{F}$, and the spectrogram attention maps $\Vec{X}_\text{map}$ obtained by doing a dot product of $\Vec{A}_\text{map}$ with the input spectrogram $\Vec{X}$.
The $\Vec{X}_\text{map}$ highlights the regions of the spectrogram that contributed to the decision made by \methodname.
We can observe that the reconstructed spectrograms $\hat{\Vec{X}}$ do not capture all the fine details of the original input spectrograms $\Vec{X}$.
They rather identify the regions and frequency bins that are more energetic.
From the observation of $\Vec{X}_\text{map}$ we can see that the detector tends to take its decisions based on mainly energetic frequency bands.
These most likely contain information related to the speech being bona fide or synthetic.
\section{Conclusions and Future Work}\label{sec:conclusions}
In this paper we propose \methodfullname~ that uses disentangled representation learning to detect synthetic speech.
Visualizing disentangled features and activation map helps to understand the working of the detector.
\methodname~ showed high performance 
compared to several existing approaches on ASVspoof2019 dataset.
In future work, we plan to replace the used variational autoencoder with a transformer neural network, as in several existing works these networks have shown better performance than variational autoencoder to learn general speech representations.
We also plan to evaluate \methodname~ on detecting noisy 
synthetic speech signals.
Lastly, currently we create unsupervised activation map, we plan to investigate supervised learning for synthetic speech localization. 

\section*{Acknowledgments}

    This material is based on research sponsored by the Defense Advanced Research Projects Agency (DARPA) and the Air Force Research Laboratory (AFRL) under agreement number FA8750-20-2-1004. 
    The U.S. Government is authorized to reproduce and distribute reprints for Governmental purposes notwithstanding any copyright notation thereon. The views and conclusions contained herein are those of the authors and should not be interpreted as necessarily representing the official policies or endorsements, either expressed or implied, of DARPA, AFRL or the U.S. Government.
    Address all correspondence to Edward J. Delp, \texttt{ace@purdue.edu}.
\vspace{-.8mm}

\bibliographystyle{IEEEtran}
\bibliography{refs}

\end{document}